\documentclass[11pt,a4paper]{article}
\usepackage[T1]{fontenc}
\usepackage[utf8]{inputenc}
\usepackage{authblk}
\usepackage{graphicx}
\usepackage{amsmath}
\usepackage{amssymb}
\usepackage[caption=false]{subfig}
\usepackage{multirow}
\usepackage{adjustbox}
\usepackage{float}
\usepackage{caption}
\usepackage{hyperref}
\usepackage{siunitx}
\usepackage{pdfpages}

\title{Author Accepted Manuscript submitted to Elsevier \\ A rigged model of the breast for preoperative surgical planning}

\author[1]{Arnaud Mazier}
\author[2]{Sophie Ribes}
\author[2]{Benjamin Gilles}
\author[1,3]{Stéphane P.A. Bordas\thanks{Corresponding author: Department of Computational Science, Université du Luxembourg, Esch-sur-Alzette, 2, avenue de l’Université, Luxembourg / Telephone: +352 621 131 048 / Fax: +352 46 66 44 35567  /  Email: stephane.bordas@me.com}}
\affil[1]{Department of Computational Science, Université du Luxembourg, Esch-sur-Alzette, Luxembourg}
\affil[2]{AnatoScope, Montpellier, France}
\affil[3]{China Medical University Hospital, China Medical University, Taichung, Taiwan}

\begin{document}
 \maketitle

\noindent Final published version available at https://doi.org/10.1007/s00366-021-01597-z. Copyright 2022. This author accepted manuscript (AAM) is made available under the CC-BY-NC-ND 4.0 license http://creativecommons.org/licenses/by-nc-nd/4.0

\newpage
\begin{abstract}
In breast surgical practice, drawing is part of the preoperative planning procedure and is essential for a successful operation. In this study, we design a pipeline to assist surgeons with patient-specific breast surgical drawings. We use a deformable torso model containing the surgical patterns to match any breast surface scan. To be compatible with surgical timing, we build an articulated model through a skinning process coupled with shape deformers to enhance a fast registration process. On one hand, the scalable bones of the skinning account for pose and morphological variations of the patients. On the other hand, pre-designed artistic blendshapes create a linear space for guaranteeing anatomical variations. Then, we apply meaningful constraints to the model to find a trade-off between precision and speed. The experiments were conducted on $7$ patients, in $2$ different poses (prone and supine) with a breast size ranging from $36$A and $42$C (US/UK bra sizing). The acquisitions were obtained using the depth camera Structure Sensor, and the breast scans were acquired in less than 1 minute. The result is a registration method converging within a few seconds ($3$ maximum), reaching a Mean Absolute Error of $2.3$ \si{mm} for mesh registration and $8.0$ \si{mm} for breast anatomical landmarks. Compared to the existing literature, our model can be personalized and does not require any database. Finally, our registered model can be used to transfer surgical reference patterns onto any patient in any position.

\end{abstract}

\section{Introduction}

In 2018 breast cancer was the second most prevalent cancer with more than $2$ million cases and an increasing incidence rate of $0.3$\% per year~\cite{Sauer2019}. Surgery remains one of the most common treatments. In $2016$, nearly one-half of patients with early-stage (Stage I or II) breast cancer underwent breast-conserving surgery~\cite{Sauer2019}.

The least invasive and traumatic operation, lumpectomy, consists of removing the breast tumor, including surrounding tissues~\cite{Riviere2018}. Before surgery, the surgeon draws surgical patterns on the patient in a preoperative position (standing). Then, the surgeon instantly operates in the intra-operative stance (supine)~\cite{Duares2019}. These patterns, also called surgical drawings, are part of the preoperative planning procedure. Most of the time, they are noticeable anatomical landmarks or a visual map that will guide the surgeon during the surgery~\cite{Duares2019}. The operation success is highly dependent on the preoperative planning and will influence the final breast shape~\cite{Rahman2011}. However, surgical drawings require experience and accuracy that can be challenging for junior surgeons, especially when no gold standard is established.

In this study, we focus on devices and methods that can satisfy two constraints:
\begin{itemize}
	\item an intuitive acquisition device capable of capturing the external shape of the patient as quickly as possible to be compatible with surgical requirements,
	\item a deformable upper-torso model embedding surgical drawings that can rapidly fit the patient scan.
\end{itemize}

To satisfy the first constraint, devices like $3\textrm{D}$ surface scanning can fulfill the surgical timing criteria~\cite{Chae2016}. Initially, these devices were expensive and limited to applications such as computer graphics. Nowadays, they are affordable and widely used for plastic surgery assistance~\cite{Tzou2014}.
To satisfy the second constraint, many techniques have been developed. Among them, non-rigid registration methods allow to reconstruct patients' breasts with high precision, template-free, and markerless by solely using a low-cost depth camera~\cite{Lacher2017, Lacher2019}. The major drawback is the execution time ranging between $1$ to $2$ hours, making the method unpracticable for our application. The finite element method relies on physics-based equations to compute realistic deformations of the model~\cite{Eiben2016} and can be improved with free form deformers for higher precision~\cite{Lee2010, Carvalho2019}. This method is computationally expensive and solving partial derivative equations requires information which is difficult to obtain (e.g. mechanical and rheological properties, boundary conditions, loading). Moreover, free form deformers require control points to improve the registration accuracy at the cost of an increased computation time. $3\textrm{D}$ morphable models ($3\textrm{DMM}$) used with landmark constraints are fast, accurate, and can support different inputs such as $3\textrm{D}$ scans or $2$D images~\cite{Kim2008, Ruiz2018}. Despite a compatible timing, the method requires anatomical landmarks to provide an initial shape and relies on a consequent database that is not compatible with our database size. 
Learned-models are widely used for their robustness to pose variation, flexibility, speed, and efficiency but require a large dataset to train the model~\cite{Casas2018, Bessa2020}. The Skinned Multi-Person Linear (SMPL) model uses corrective body-shape and pose-dependent shape learned from thousands of $3\textrm{D}$ body scans to provide a real-time, partially open-source, and realistic articulated model~\cite{Pons-moll2015}. Despite viable medical applications~\cite{Hesse}, the model relies on an unadapted learning database made of women wearing bras. It makes it impossible for us to capture anatomical details that are only visible on bare breasts. To the authors' knowledge, there is no open-source database of $3\textrm{D}$ scans of bare breasts available.

In this paper, we propose a flexible model allowing to fit a set of patients in different surgical positions, without any previous training phase or mesh pre-treatment.
To ensure a quasi-instantaneous patient fitting, we developed a simple articulated model made of virtual bones to permit pose modifications, namely skinning~\cite{James2005, Baran2007, Jacobson2011, Le2014}. This model is made more adaptive by allowing bone scalability to cater for variations in body morphologies. Moreover, we used a simple linear model of body shapes to account for morphological variations in breast shape, also known as blendshapes~\cite{Parke1972, Joshi2005, Beacco2012, Lewis2014}. 
In this work, we assumed that an accurate registration enables a mapping of the model's surgical drawing to any patient scan. Indeed, the model gathers anatomical features and surgical patterns made by a senior surgeon, thus providing a gold standard for inexperienced surgeons.
The final aim of this study is to provide a pipeline to assist junior surgeons with preoperative breast surgical drawings for lumpectomy surgery, including, a patient-specific model ready for biomechanical simulations.

\section{Method}

\subsection{Participants}

The study was performed on $7$ women, in preoperative and intra-operative positions. The preoperative configuration corresponds to the stand-up position where the surgeon draws the surgical patterns on the patient. The intra-operative stance corresponds to the supine position (the patient lays on the back) for the surgery. Their ages varied between $47$ and $69$ years and the surgeon manually measured their breast size from $36$A to $42$C in US/UK bra sizing~\cite{Zheng2006}. We noticed one notable case of breast asymmetry (one breast had a different size or volume to the other one). All the patients were diagnosed with breast cancer and chose lumpectomy as the most suitable treatment, under medical recommendation. We obtained consent from the $7$ patients for the study and a favorable opinion from the “Comité Local d’éthique Recherche”\footnote{ Obtained on the $07/16/2017$ under the label $2017\_\texttt{CLER-MTP}\_07\texttt{-}04$. The study was declared in the registry of the CNIL (MR$003$) under the name of the "Centre Hospitalier Universitaire" (CHU) of Montpellier and registered on the ClinicalTrials website \url{https://www.clinicaltrials.gov/ct2/show/study/NCT03214419}.}

\subsection{Instruments}

To obtain a $3\textrm{D}$ surface mesh of the patient we used the depth camera Structure Sensor $3\textrm{D}$ Scanning by Occipital. This scanning device suffers from a high noise-sensitivity and the possible creation of spurious gaps within the mesh. However, it satisfies our criteria in terms of rapid acquisition time, a user-friendly interface, and reasonable price.

\subsection{Procedures}

After sketching the preoperative surgical drawing, the surgeon used the device mounted on a digital tablet to scan the upper-body of the patient. To obtain global features of the patient, the surgeon circled around the target at approximately $1$ \si{m}. Otherwise, to capture more complex shapes like the breast fold, the surgeon should be closer to the target (around $30$ \si{cm}). The device aimed at the patient for $20$ seconds to $1$ minute, depending on the desired resolution. The surgeon performed the 14 scans without any previous training. The result was a surface mesh of the patient obtained in less than $1$ minute, including the scan and the reconstruction.

\subsection{Rigging}

Each bone of the rig is considered as a rigid-scalable object, meaning that each bone will have $9$ absolute degrees of freedom (DOFs) ($3$ rotations, $3$ translations, and $3$ for the scale). The relative motion of the bones is constrained by the joints. We computed local rotations and translations of the joints by computing relative transformations at a rigid point attached to the scalable bones, represented by the spheres in Figure $1a$. Constraints were then applied through Lagrange multipliers on both local rotations and translations~\cite{Tournier2017}. Quaternions were used to define the axis angle of every bone. From this, we computed a local rotation matrix. 

\begin{figure}[ht]
	\includegraphics[width=\textwidth]{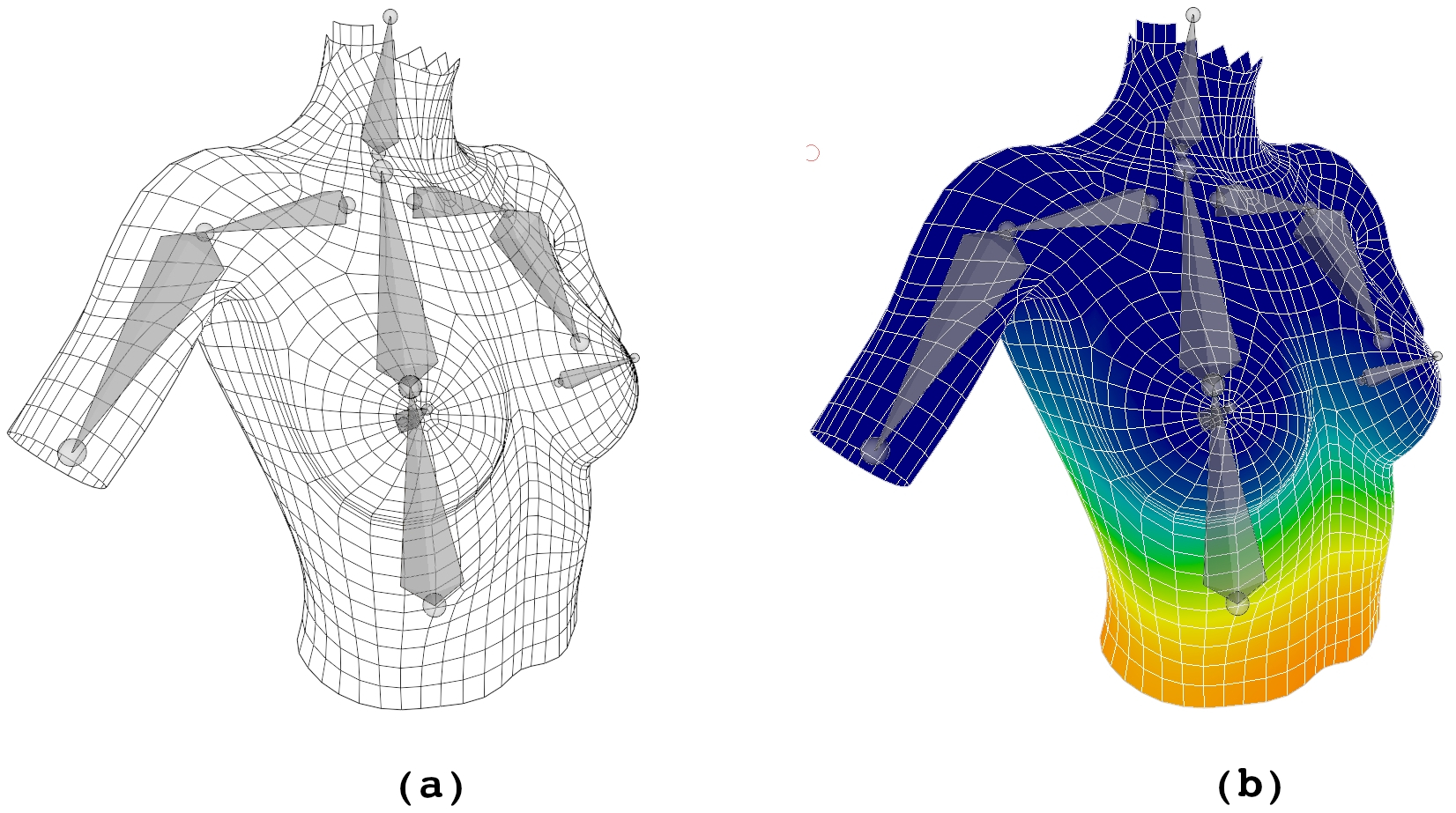}
	\caption{(a) Template mesh skin (in wire-frame) and virtual bones (in grey). (b) Blend weight colormap associated to the lower bone (red for a weight of 1 and blue for a weight of 0).}
	\label{fig:1}
\end{figure}

\subsection{Skinning}
We used blend weights to adjust the influence of each skeletal bone on the skin. In Figure $1b$, we displayed the colormap of the blend weights associated with the lower bone. The red regions are rigidly attached to the bone and follow the motion of the bone. Conversely, blue areas are not affected by the movements of the bone. The blend weight matrix can be manually given by the user or can be automatically calculated by $3\textrm{D}$ modeling software such as Blender \footnote{https://www.blender.org/}.

In this study, the model was made of $K = 9$ bones and $N = 2200$ vertices as shown in Figure $1a$. We used the following notations: $R_{j}({q})$ the rotation matrix of the $j$-th bone obtained from the quaternion rotation, $T_{j}$ the associated vector bone offset and $S_{j}$ the scale matrix with a scale coefficient for each direction. As a result, with a given set of $K$ bones, bones rotation $R(q) = [R_1(q),...,R_K(q)]$, bone offset $T = [T_1,...,T_K]$, bone scale $S = [S_1,...,S_K]$. By calling the elements of the blend weight matrix $w_{j, i}$ ($w \in {\rm I\!R}^{K \times N}$) and the rest template mesh vertices $v_i$ ($v \in {\rm I\!R}^{N \times 1}$). The computed vertices ${M}_{i}$ are given by the equation of the Linear Blend Skinning (LBS):

\begin{equation}
\label{eq:skinning}
M_{i}({R(q),S,T})=\sum_{j}^{K} w_{j, i} ({R}_{j}(q){S}_{j} {v}_{i}^T + {T}_{j}),
\end{equation}

\subsection{Blendshapes} 
For our model, an artist created $A = 55$ blendshapes affecting global features such as shoulders or belly size as well as more local ones establishsuch as nipples and aureole shapes. 
Let $v$ denotes the vertex positions of the template mesh, $B$ the blendshape function computing the deformed vertex positions and ${b}_{k}$ the $k$-th shape displacement matrix, with $v, B, {b}_{k} \in {\rm I\!R}^{N \times 3}$. Blendshape displacements provide a set of basis vectors that define a linear space which is used to generate a vector space, onto which the patient’s shape is projected. 

\begin{equation}
\label{eq:blendshape}
B(\alpha_{k}) =v+\sum_{k}^{A} \alpha_{k}{b}_{k},
\end{equation}
where $A$ is the number of blendshapes and $\alpha_{k} \in \mathbb{R}^{A}$ the linear blendshape weights affected to each $k$-th blendshape. To ensure convexity and invariance for rotation and translation, these weights should fulfil the following conditions: $\sum_{i=1}^{n} \alpha_{k}=1$ and $\alpha_{k} > 0,\forall k \in [ 1,n ] $. These 55 blending weights ($\alpha_{k}$) can be used as DOFs for our deformable model, we call $\alpha$ the vector storing the $\alpha_{k}$ values.

By combining different body shapes, the model can cover a large deformation space to fit several morphologies (Figure $2$) and can be easily enriched by adding more blendshapes.

\begin{figure}[ht]
	\includegraphics[width=\textwidth]{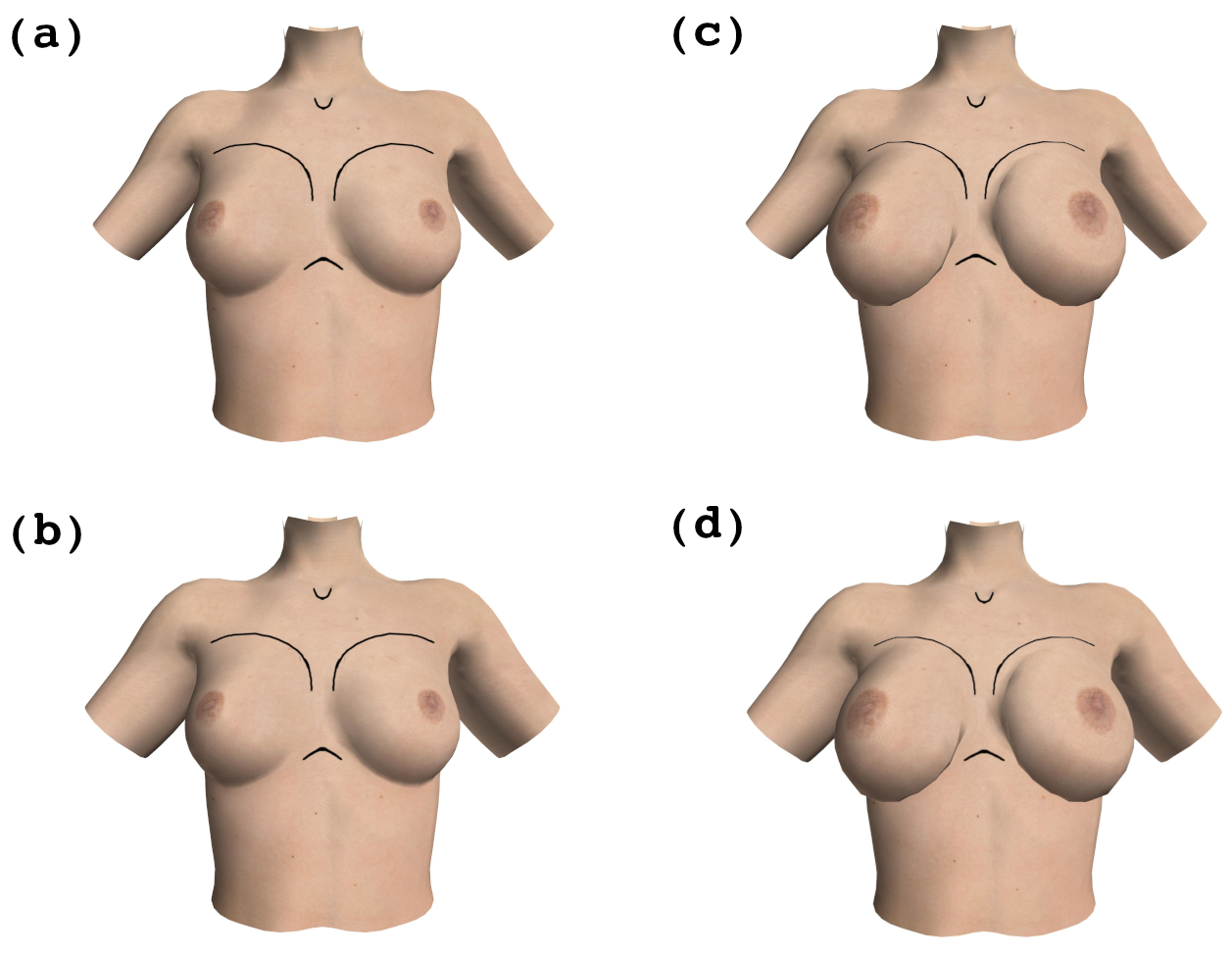}
	\caption{(a) Template mesh. (b) First blendshape affecting the breast size ($\alpha$0b0 with $\alpha$0=1). (c) Second blendshape affecting the arm size ($\alpha$1b1 with $\alpha$1=1). (d) First and second blendshapes activated ($\alpha$0b0+$\alpha$1b1 with $\alpha$0=$\alpha$1=1).}
	\label{fig:2}
\end{figure}

\subsection{Final model}
In the final model, we combined the effect of blendshapes mixed with the scalable bones from the skinning. By inserting equation \ref{eq:blendshape} into \ref{eq:skinning}, we obtained:
\begin{equation}
M_{i}({R(q),S,\alpha,T}) =\sum_{j}^{K} w_{j, i} ({R}_{j}(q){S}_{j} B_{i}^T(\alpha) + {T}_{j}),
\end{equation}

To avoid distortion of the model and to regularize the energy minimization we defined $3$ additional energy terms. 
A scale energy ${E}_{S}$ to penalize the scale matrix of each bone towards its original scale (with $\mathbb{I}_{3}$ the order $3$ identity matrix). 
A blendshape energy ${E}_{BS}$ to regularize the blendshape weights towards 0 and a joint energy ${E}_{J}$ to regularize the translations and the relative rotations of each joint to their initial configurations, respectively ${T}_{j}^{*}$ and ${R}_{j}(q^{*})$ for the $j^{th}$ bone. With $\parallel \bullet \parallel$ the Euclidean norm, we obtained the following equations:
\begin{equation}
{E}_{S} ({S}) = \sum_{j}^{K} \parallel S_{j} - \mathbb{I}_{3}\parallel ^{2},
\end{equation}
\begin{equation}
{E}_{BS} ({\alpha}) = \parallel \alpha \parallel,
\end{equation}
\begin{equation}
{E}_{J}({R(q),T}) = \sum_{j}^{K} \parallel \log ({R}_{j}(q^{*}){R}_{j}^{-1}(q)) \parallel ^{2} + \parallel {T}_{j} - {T}_{j}^{*} \parallel ^{2}.
\end{equation}

\subsection{Registration}
\label{section:registration}
We chose the Simulation Open Framework Architecture (SOFA ~\cite{faure:hal-00681539}) to perform the registration of our model to the $7$ patients in preoperative and intra-operative stance. To fit the skin of the model to the scan of the patient, we chose to minimize the closest-point distances such as in the Iterative Closest Point (ICP) algorithm~\cite{Besl1992}. More precisely, we used an octree structure to find the closest vertex on the scan from source vertices. Then, we projected the source point onto the closest primitive (triangle, edge, or point) around each closest vertex. This allows for a more accurate registration rather than using point-to-point distances and differentiating \ref{eq:closest} with respect to $M_i$ can be easily done by computing the normal vectors of the triangles of the scan. As the scanning process depends on the surgeon, some scans can be incomplete with only the front view and artifacts, as shown in Figure $3a$. To improve the robustness of the algorithm against noise and local solutions, filters to reject outliers were added (distance and normal vector threshold). 

Based on correspondences established at each iteration, we identified the best set of parameters that minimized the distance from the scan to the deformed model. We defined a data energy term that penalizes the squared Euclidean distance between the model vertices $M_i$ and the target mesh vertices $V_j$ according to the closest point algorithm:
\begin{equation}
\label{eq:closest}
    {E}_{D} ({R(q),S,\alpha,T}) = \sum_{(i,j)}^{N_{CP}} \parallel d_{CP} \parallel^{2} = \sum_{(i,j)}^{N_{CP}} \parallel M_{i}({R(q),S,\alpha,T}) - {V}_{j} \parallel^{2},
\end{equation}
where $N_{CP}$ represents the index pairs found by the closest point algorithm and $d_{CP}$ the distance between $V_j$ the closest point from $M_i$ on the set of target mesh triangles. The total energy to minimize is:
\begin{equation}
   E_{\mathrm{tot}}(R(q),S,\alpha,T) = \dfrac{1}{\lambda_{D}}E_{D} + \dfrac{1}{\lambda_{S}}E_{S} + \dfrac{1}{\lambda_{BS}}E_{BS} + \dfrac{1}{\lambda_{J}}E_{J}
\end{equation}
with $\lambda_{D} = 10e^{-3} ,\lambda_{S} = 10e^{-2}, \lambda_{BS} = 10e^{-3}, \lambda_{J} = 10e^{-2}$ set empirically. We used a regularized Newton algorithm and stop the minimization when we reached our convergence threshold based on the distance compared to the last iteration. 

As described in~\cite{Tournier2017}, we iteratively minimized equation \ref{eq:closest} using an implicit integration of Newton's equation, using a compliant formulation to handle both stiff constraints (joint translations and rotations) and elastic terms in a stable manner. The solver finds a compromise between minimizing the distance to the data and the distortion of the reference model.

According to~\cite{Farinella2006}, we manually added $12$ anatomical landmarks on all scans and on the model in preoperative and intra-operative stance (respectively Figure $3a$ and $3b$). Among the $12$ landmarks, $2$ have no symmetry along the sternum from top to down: the sternal notch and the xiphoid. Then, $6$ symmetric landmarks (right and left) from top to bottom: the acromial extremity of the clavicle, the mid-axillary point, the pectoralis the insertion in the arm, the nipple, and the lowest breast point with respect to the vertical body axis. These landmarks were chosen for their easy reproducibility, as validation criterion for the registration, and to possibly strengthen our registration.

Our registration approach offers the possibility to combine automatic vertex matching (closest-point algorithm) with manual vertex matching (landmarks).
By enabling the surgeon to interactively select these landmarks, we can add a landmark energy term to penalize the squared distance between the model landmarks and the scans landmarks. The landmark energy ${E}_{L}$ expression is similar to equation \ref{eq:closest}, but we replaced the closest-points index pairs with the corresponding landmarks index pairs.

\begin{equation}
\label{eq:finale}
E_{\mathrm{fin}}(R(q),S,\alpha,T) =  E_{\mathrm{tot}} + \dfrac{1}{\lambda_{L}}{E}_{L},
\end{equation}
with $\lambda_{L} = 10e^{-4}$ also set empirically.

Finally, we drew on the model the surgical drawing of a lumpectomy. They were made of $4$ shapes: $2$ arrows delineating the upper and lower sternum, $2$ curves delimiting the right and left breast (Figure $3b$).

\begin{figure}[ht]
	\includegraphics[width=\textwidth]{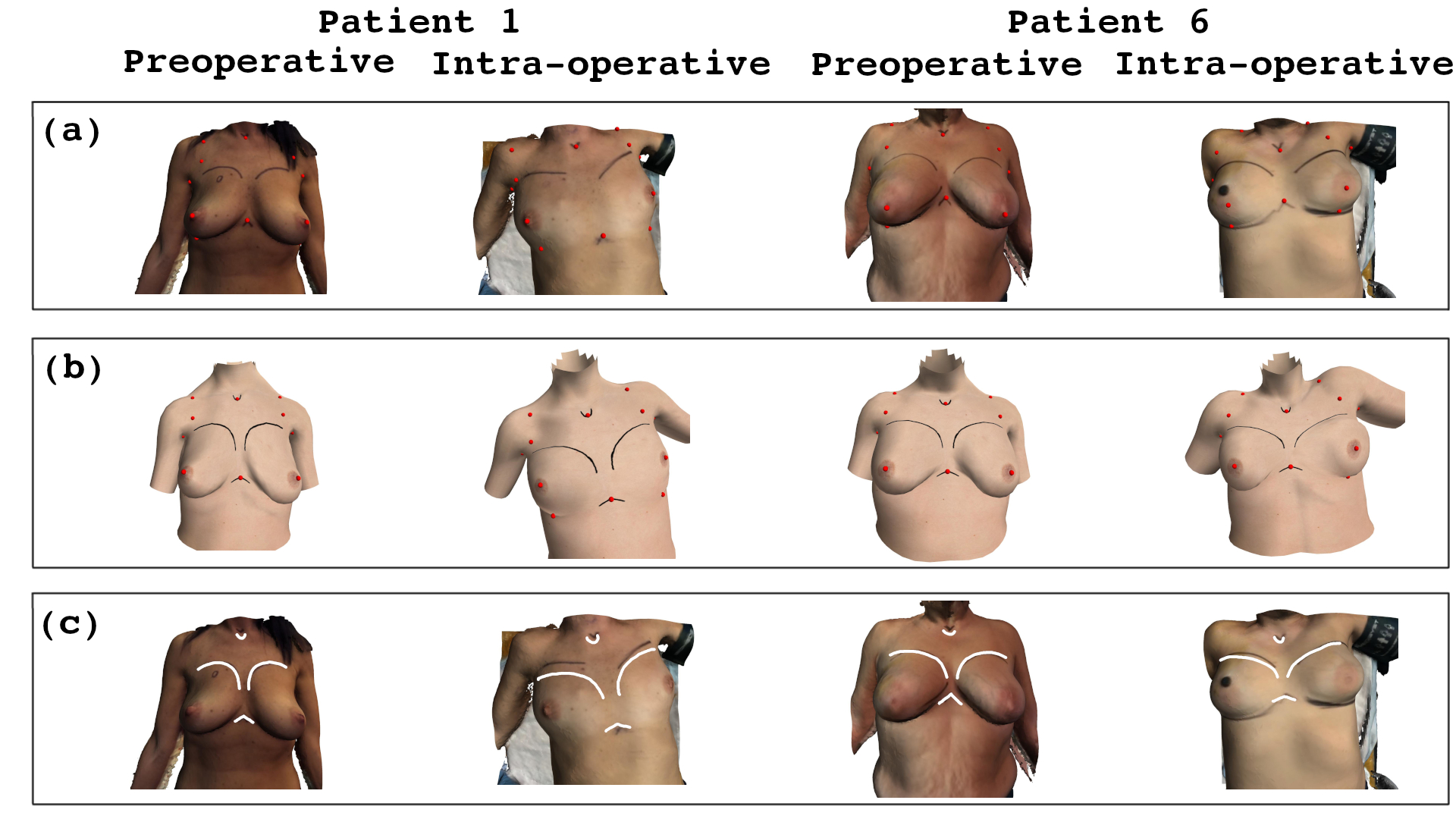}
	\caption{Comparison of surgical pattern drew by the surgeon and guessed pattern by the model. (a) Scan of patients 1 and 6 in preoperative and intra-operative stance. The red spheres represent the 12 landmarks described in [Farinella et al., 2006] and the black shapes are the surgical drawing made by the surgeon on the patient. (b) Model registered on the preoperative and intra-operative patient scan. The red spheres represent the 12 landmarks described in [Farinella et al., 2006] and the black shapes are the surgical drawing made by the surgeon on our model. (c) Superimposition of the surgical drawing made by the surgeon on the patients (black) and our surgical drawing estimation of our model (white).}
	\label{fig:3}
\end{figure}

\subsection{Statistics}
To assess the quality of the registrations, we used the Python packages NumPy~\cite{Haris2020} and SciPy~\cite{Virtanen2020} to calculate statistical quantities.
For each of the $14$ registrations, we recorded the time needed to reach the convergence criteria of the solver. Then, we evaluated the surface registration error by calculating the final distances between the vertices of the model and the vertices of the scans ($d_{CP}$). We did the same for the landmarks registration error by calculating the distances between the landmarks of the model and the landmarks of the scans ($d_{L}$). For each distance vector, in order to avoid signed distances, we computed the $\textrm{MAE}=\frac{\sum_{i=1}^{n}\left|d_{i}\right|}{n}$, where $n = N_{CP}$ and $d = d_{CP}$ for the surface MAE, whereas $n = 12$ and $d = d_{L}$ for the landmarks MAE. Finally, we computed the standard deviation, minimum and maximum values of the absolute distances.

\section{Results}

\begin{table}[ht]
	\centering
	\begin{adjustbox}{width=1\textwidth}
		\begin{tabular}{c | c | c | c | c | c | c | c | c }
			\multicolumn{9}{c}{\textbf{Preoperative configuration}} \\
			\hline
			\multirow{2}{*}{\textbf{Patient}} &
			\multirow{2}{*}{\textbf{Time} (s)} &
			\multicolumn{3}{c|}{\textbf{Surface distance error}} & 
			\multicolumn{4}{c}{\textbf{Landmarks distance error}} \\ 
			&&MAE (mm)&SD (mm)&Max (mm)&MAE (mm)&SD (mm)&Max (mm)&Min (mm)\\ \hline
			0 & 1.98 & 2.31 & 3.24 & 23.7 & 8.24 & 3.42 & 13.8 & 3.74  \\ \hline
			1 & 2.17 & 2.22 & 4.06 & 45.6 & 8.70 & 4.99 & 16.2 & 4.68 \\ \hline
			2 & 2.11 & 2.58 & 2.94 & 35.9 & 11.03 & 5.88 & 25.0 & 2.34 \\ \hline
			3 & 1.32 & 2.23 & 2.15 & 14.5 & 8.98 & 3.97 & 16.2 & 1.61  \\ \hline
			4 & 1.38 & 2.43 & 2.88 & 25.0 & 8.17 & 6.54 & 20.7 & 0.94 \\ \hline
			5 & 2.23 & 2.23 & 3.39 & 23.3 & 7.22 & 3.46 & 11.9 & 1.21 \\ \hline
			6 & 1.79 & 2.87 & 3.71 & 51.5 & 6.89 & 5.72 & 22.9 & 3.53 \\ \hline
			\textbf{Mean} & \textbf{1.85} & \textbf{2.41} & \textbf{3.19} & \textbf{31.4} & \textbf{8.46} & \textbf{4.85} & \textbf{18.1} & \textbf{1.98} \\
			\\
			\multicolumn{9}{c}{\textbf{Intra-operative configuration}} \\
			\hline
			\multirow{2}{*}{\textbf{Patient}} &
			\multirow{2}{*}{\textbf{Time} (s)} &
			\multicolumn{3}{c|}{\textbf{Surface distance error}} & 
			\multicolumn{4}{c}{\textbf{Landmarks distance error}} \\ 
			&&MAE (mm)&SD (mm)&Max (mm)&MAE (mm)&SD (mm)&Max (mm)&Min (mm)\\ \hline
			0 & 2.01 & 2.39 & 3.48 & 31.9 & 9.36 & 3.68 & 14.9 & 3.40 \\ \hline
			1 & 1.90 & 2.11 & 3.00 & 17.7 & 7.96 & 4.13 & 15.2 & 1.54 \\ \hline
			2 & 1.99 & 2.14 & 3.13 & 20.1 & 7.23 & 3.74 & 14.0 & 1.02 \\ \hline
			3 & 2.54 & 2.33 & 2.78 & 19.6 & 3.95 & 3.40 & 12.3 & 0.37 \\ \hline
			4 & 2.14 & 1.85 & 2.78 & 24.3 & 7.32 & 3.44 & 11.2 & 0.21 \\ \hline
			5 & 3.01 & 2.44 & 2.86 & 20.4 & 7.03 & 4.91 & 15.8 & 2.84  \\ \hline
			6 & 1.80 & 2.70 & 3.43 & 30.6 & 10.39 & 4.36 & 19.6 & 2.85  \\ \hline
			\textbf{Mean} & \textbf{2.20} & \textbf{2.28} & \textbf{3.06} & \textbf{23.5} & \textbf{7.61} & \textbf{3.95} & \textbf{14.7} & \textbf{1.75}  
		\end{tabular}
	\end{adjustbox}
	\caption{Registration statistics errors for preoperative and
		intra-operative stance.}
	\label{tab:tab1}
\end{table}

Table $1$ shows the result of the $14$ registrations and the Figure $3b$ the visual result of the registration of our model on $2$ patients. On average, the registration process took a bit more than $2$ seconds. The surface MAE for preoperative and intra-operative were respectively $2.41$ and $2.28$ \si{mm}, while the results were higher for the landmarks MAE, $8.46$ and $7.41$ \si{mm}. In Figure $3c$, we also display the surface scan of the patient with the prediction of the surgical drawing of our model (in white) compared to the surgeon's drawing (in black).

Furthermore, we demonstrated the impact of the landmarks on the registration process in Figure $4$. Indeed, without using the landmarks in the registration, we achieved a MAE of $2.01$ \si{cm} between the landmarks of the model and the scans, which was too large. By taking the $12$ landmarks into account, we improved the landmarks MAE to $8.03$ \si{mm} without increasing the surface MAE.

Moreover, we studied the effect of blendshape numbers on the performances of the model. As shown in Figures $5$ and $6$, for all patients and the $2$ configurations, increasing the number of blendshapes decreased the surface MAE. Hence, we observed a non-constant loss that seemed to be triggered after reaching a certain number which was not the same for all patients. 

Finally, we conducted a sensitivity analysis of the model parameters in both configurations. As we empirically chose the values of $\lambda_{D}, \lambda_{S}, \lambda_{BS}, \lambda_{J}$ and $\lambda_{L}$, we investigated the impact of these parameters on the surface and landmarks MAE. To that extent, we modified one particular parameter while fixing all other parameters to their original value and calculated the mean MAE for all patients. As mentioned in section \ref{section:registration}, these parameters represent the impact of specific energies on the entire system (equation \ref{eq:finale}). The results are displayed in preoperative configuration, in Figures $7$ and $8$, but the curves are similar in intra-operative stance. 

\begin{figure}[ht]
	\includegraphics[width=\textwidth]{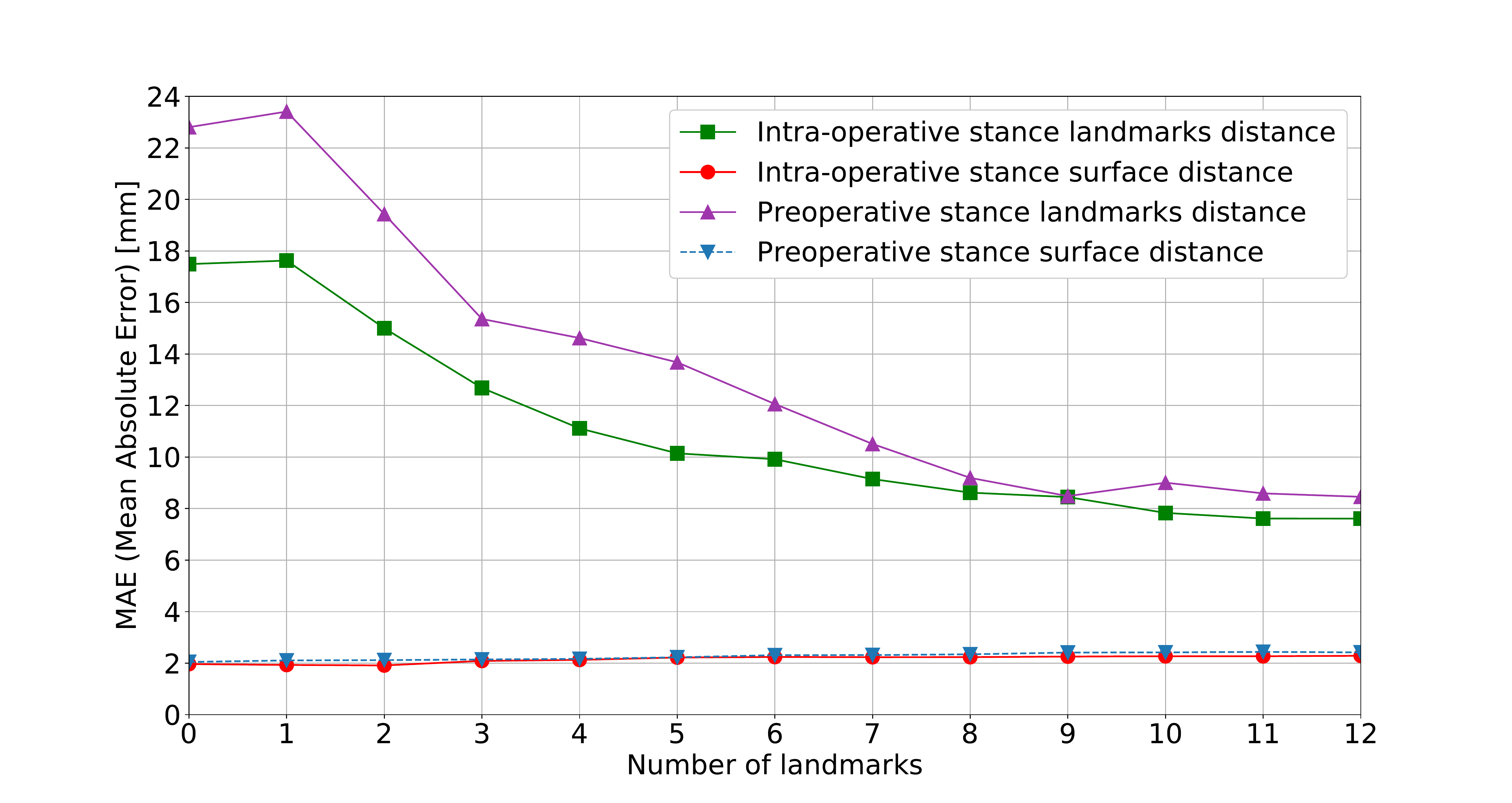}
	\caption{Mean Absolute Error (MAE) by increasing the number of landmarks in equation 9 according to [Farinella et al., 2006].}
	\label{fig:4}
\end{figure}

\begin{figure}[ht]
	\includegraphics[width=\textwidth]{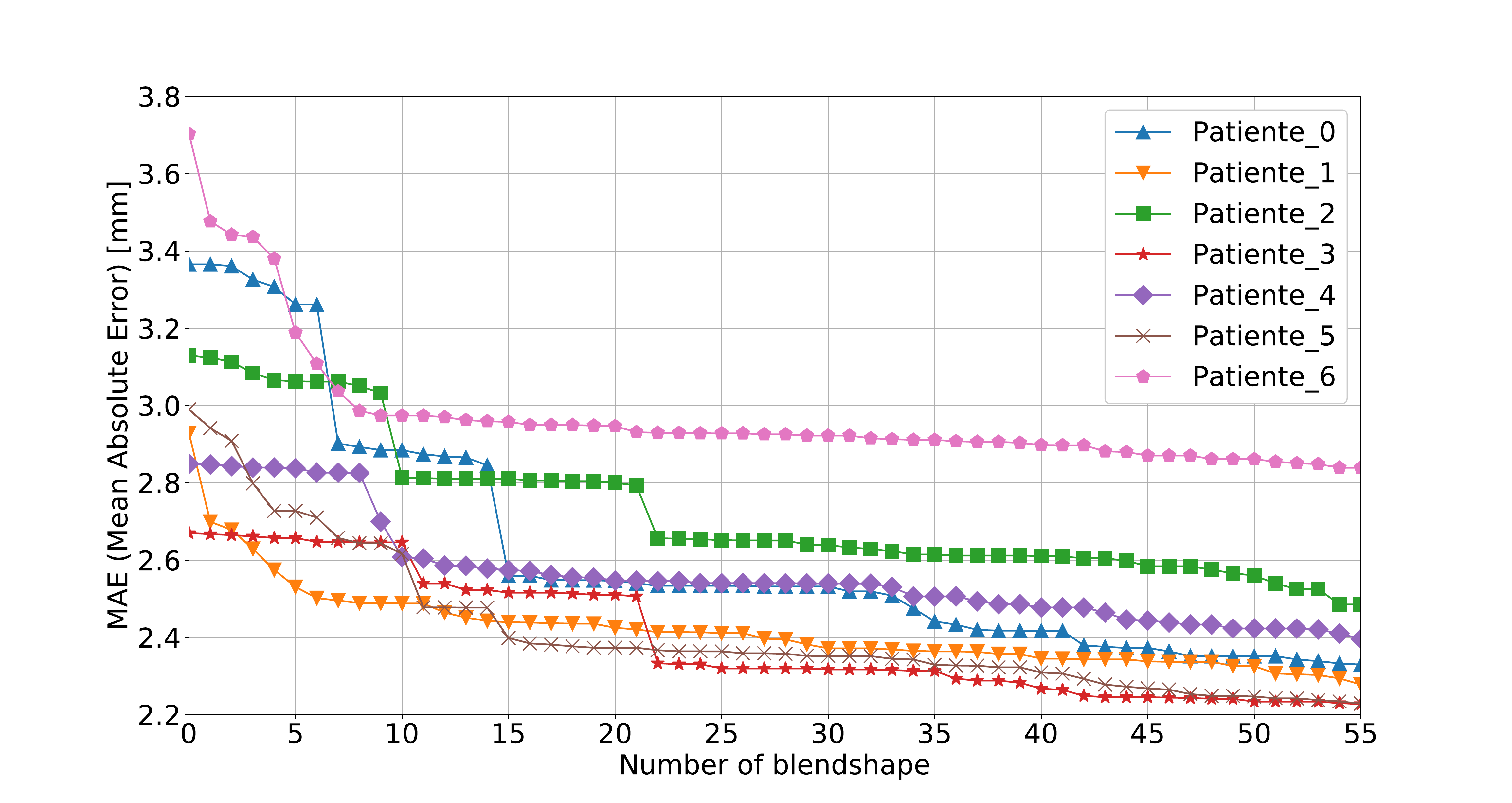}
	\caption{Mean Absolute Error of the surface (surface MAE) by increasing the number of blendshape in preoperative stance.}
	\label{fig:5}
\end{figure}

\begin{figure}[ht]
	\includegraphics[width=\textwidth]{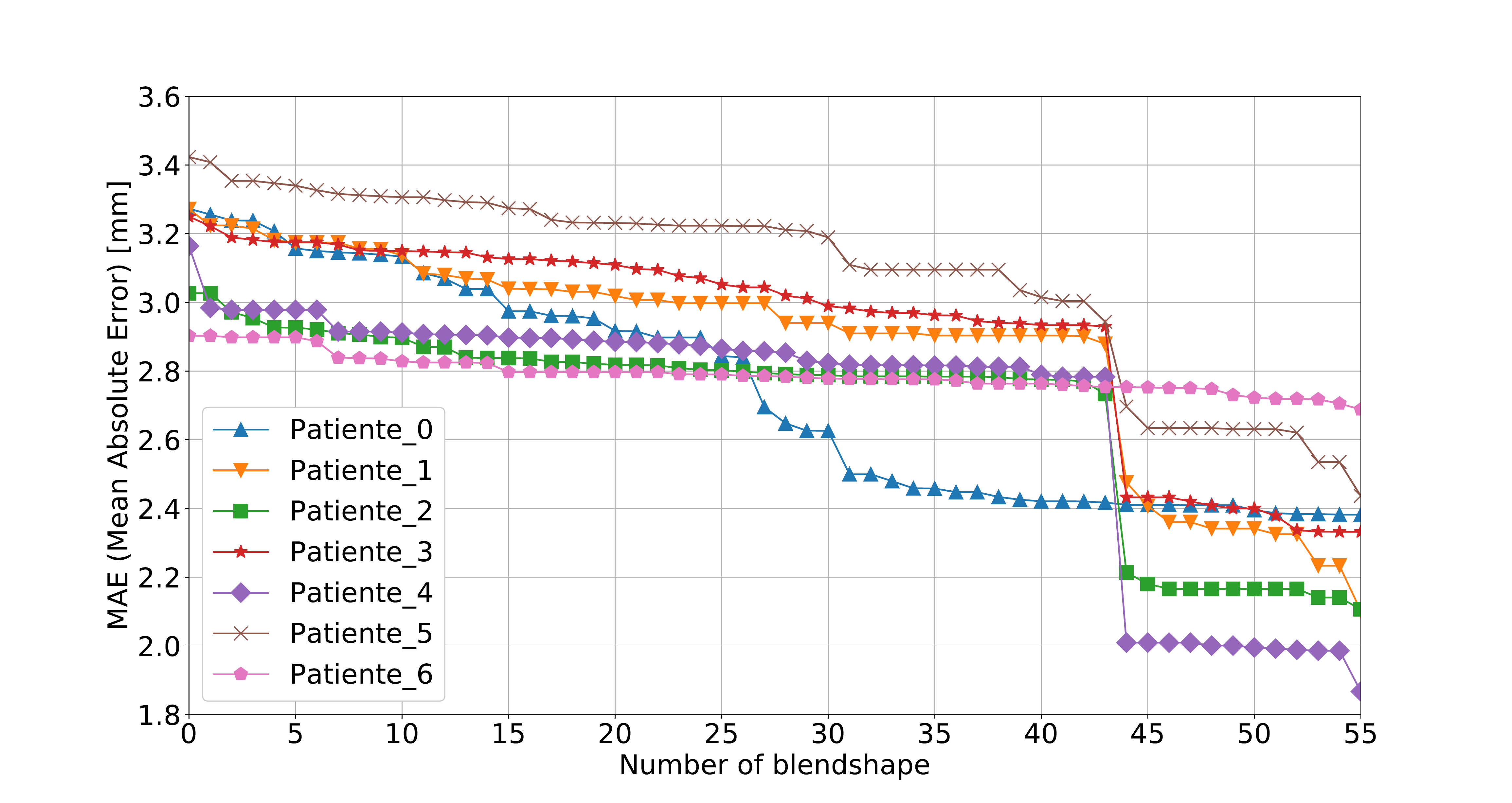}
	\caption{Mean Absolute Error of the surface (surface MAE) by increasing the number of blendshape in intra-operative stance.}
	\label{fig:6}
\end{figure}

\begin{figure}[ht]
	\includegraphics[width=\textwidth]{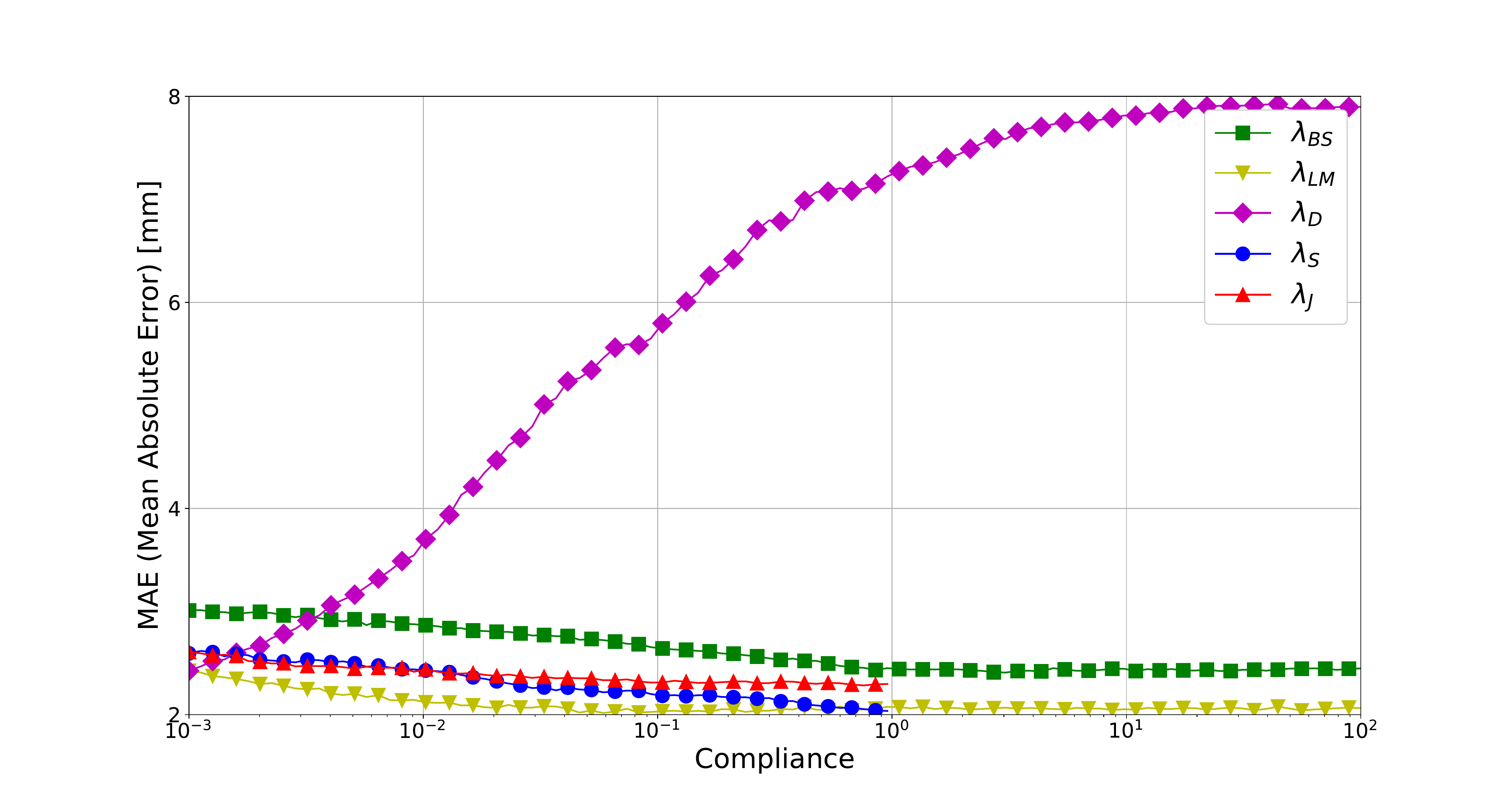}
	\caption{Sensibility analysis of the surface Mean Absolute Error (surface MAE) of the model in preoperative stance.}
	\label{fig:7}
\end{figure}

\begin{figure}[ht]
	\includegraphics[width=\textwidth]{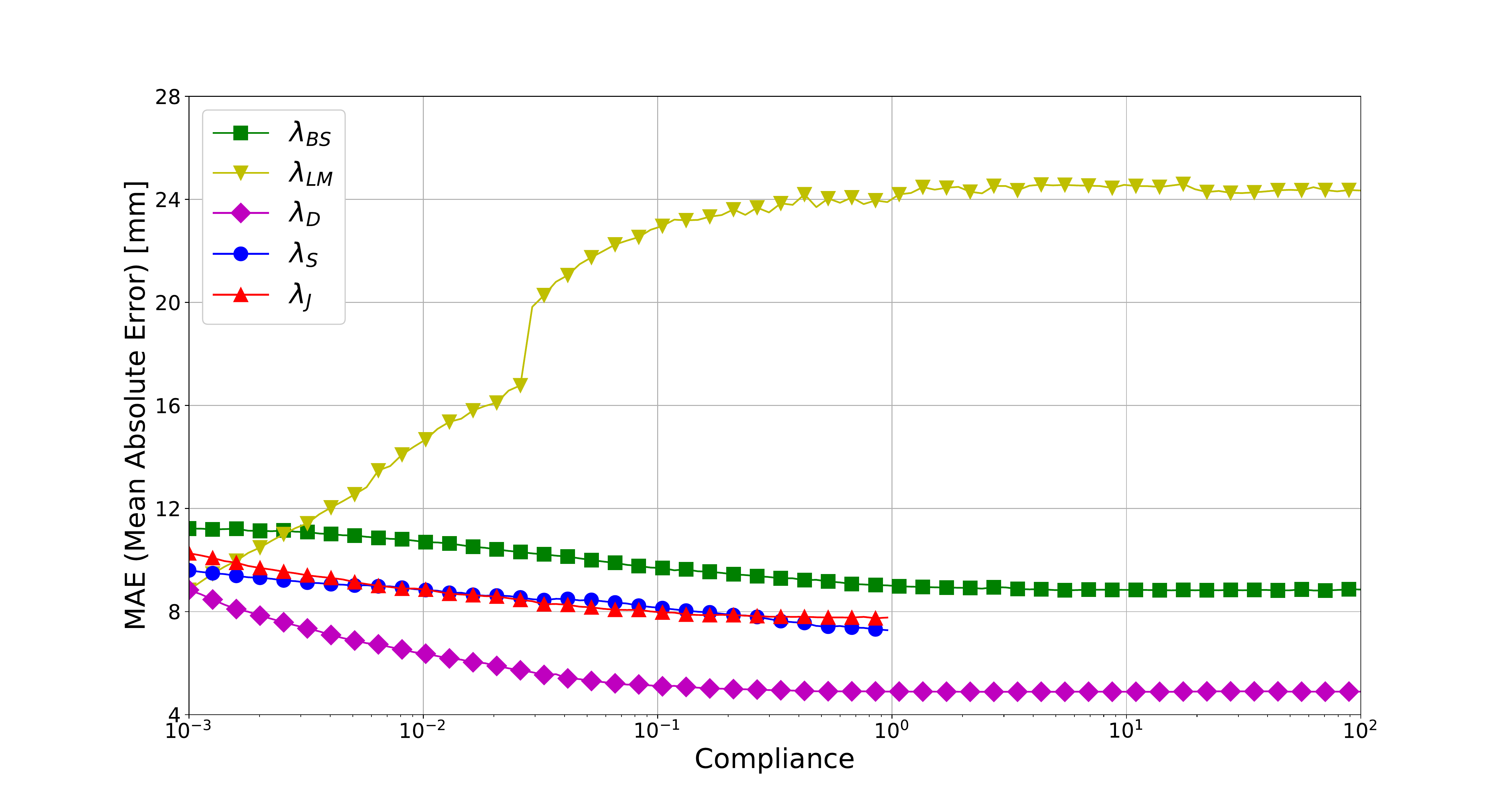}
	\caption{Sensibility analysis of the surface Mean Absolute Error (surface MAE) of the model in preoperative stance.}
	\label{fig:8}
\end{figure}

\section{Discussion}

The research presented shows that the registration is efficient and independent of the pose or the morphology. Indeed, we observed a slight difference in surface and landmarks MAE between the preoperative and the intra-operative position. Indeed, we noticed a fair matching between the surgical patterns drew by the surgeon and the prediction of our model (Figure $3$c). These results support our hypothesis that an accurate registration (low MAE) leads to an overlapping of the surgical drawings of the model (made by the senior surgeon) and of the scan, becoming a gold standard for inexperienced surgeons. The registration results obtained in this study are congruent with those present in the literature.~\cite{Ruiz2018} used $3\textrm{D}$ Morphable Models of the breast to fit two possible inputs: 2D photos and $3\textrm{D}$ scans. For $310$ $3\textrm{D}$ scans, an average distance error of $2.36$ \si{mm} in $3.15$ \si{s} was obtained with a standard deviation of $0.18$ \si{mm}. While these results are similar in magnitude to those of the present study, ~\cite{Ruiz2018} produced a lower standard deviation error (probably due to the high number of scans) but did not treat articulated movements of the patients. \cite{Carvalho2019} used Free Form Deformers to register MRI images to surface scan (obtained with the Kinect camera) with a Euclidean error lower than $1$ \si{mm} a Haussdorf distance of $4.34$ \si{mm}. Hence,~\cite{Carvalho2019} model does not support pose variation and the authors did not communicate about the computational time.

We showed that increasing the number of landmarks decreases the MAE to enhance the registration accuracy. This improvement is expected but highlights, in some cases, an inaccurate matching of the closest point algorithm. Indeed, some landmarks such as the nipples are geometrically significant on the mesh and can be detected by the automatic algorithm; others like the sternum are only textured information and struggle to be detected by the closest point procedure. Manually adding the landmarks takes less than $30$ seconds and produces a more accurate registration using less computational time.
	
By expanding the number of blendshapes, the deformation space of the model is also expanded. As a result, the model can fit more complex body shapes. Hence, determining the right amount of blendshapes can be delicate and a compromise has to be found between accuracy and registration time. On one hand, increasing the number of blendshapes augments the number of DOFs leading to higher registration time. On the other hand, one blendshape can be meaningful for one particular morphology but meaningless for another one. For example, in Figure $6$, blendshape $44$ had a significant impact on the majority of the patients but not for \emph{Patient 0} and \emph{Patient 6}. This result is also highlighted in~\cite{Pons-moll2015} where they increased the number of blendshapes from $1$ to $300$ and decreased the mean absolute error from $6.6$ \si{mm} to $3.1$ \si{mm}. In our case, for the proposed application, $55$ blendshapes are a reasonable balance with a mean accuracy lower than $3$ \si{mm} for the surface registration and an execution time below to $3$ \si{s}.

Finally, we demonstrated the robustness of the model through a sensitivity analysis. Hence, we highlighted once again, the inaccurate matching of the closest-point algorithm when we increased the impact of $E_{D}$ (decreasing the value of $\lambda_{D}$ to improve the surface MAE) which resulted in a decrease of the landmark MAE. Furthermore, we showed a weak impact of $\lambda_{S}, \lambda_{BS}, \lambda_{J}$ (which prevent model distortion) as reducing their values only force the model to its initial configuration. Conversely, selecting a value upper than $10$ for $\lambda_{S}$ and $\lambda_{J}$ led to singularities as we did not constrain the scale and the joints rotation and translation of the model. It is well-known that LBS can suffer from the "candy-wrap" effect, i.e., the mesh loses volume when joints rotations are too important. But this effect can be alleviated by using Dual Quaternion Skinning~\cite{Kavan2008} or Implicit Skinning~\cite{Vaillant2013}. So far, we did not observe such behaviors in our application. Finally, the model is flexible and can be easily modified by tuning the regularization parameters or by adding new blendshapes, landmarks, and bones.

Several limitations are acknowledged. One major limitation is the low number of participants ($7$) leading to a weak anatomical variability and a difficult validation. Through the sensitivity analysis, we showed the impact of the parameters ($\lambda$) that penalize the different energy terms. Hence, setting these parameters is not simple as they are just penalization terms and it can be hard to find the optimal ones for each model. Furthermore, we manually created our blendshapes to reproduce the soft tissue deformations that require knowledge and experience. Despite overcoming the problem of a database, generating unrepresentative blendshapes can add unnecessary DOFs to the model.
 
Suggestions for future work include the automatic detection of the landmarks, for instance, by using the scan textures. These are for the moment identified manually by the surgeon which is an advantage for flexibility but a drawback for automation. Thus, to reduce the computational time, we used a coarse mesh. But the refinement of the mesh can lead to better registration with a slight increase of the registration time. Another step forward to improve the validation would be to have access to a larger population in order to create a statistical database. This could be done by adding a Bayesian regularisation term~\cite{Rappel2016} to the energy minimization term. Finally, the registration provides a patient-specific mesh ready for biomechanical simulations. Hence, this work can be the basis for pre to the intra-operative mapping of tumors using the finite element method.

\section{Conclusion}
In the present study, we registered a simplified breast-model on $7$ patients in the preoperative and intra-operative configurations. We showed that Linear Blend Skinning was a good approximation of the bones joint motion and allowed the model to fit different poses. Morphological differences as well as soft-tissue deformation induced by different poses, could be modeled by using blendshapes. Our application was mainly focused on lumpectomy but the methodology is general enough to be compatible with other surgical patterns such as mastectomy or mammary reduction. The method could even be generalized to other body parts but would require creating a new rig, mesh, and blendshapes.
We showed a concrete clinical application of breast patient-specific modeling for preoperative surgery drawing based on real data.
In the end, our model fits a scan in less than $3$ seconds, is robust to noise, incomplete data, posture, and morphological variations of patients.

\section*{Acknowledgements}
This study was supported by the European Union’s Horizon 2020 research and innovation program under grant agreement No 811099 and the Marie Sklodowska-Curie grant agreement No. 764644. The medical images used in the present study were obtained from Hopital Arnaud de Villeneuve, Département de Gynécologie Obstétrique in collaboration with Dr. Gauthier Rathat. The authors would like to thank Dr. Jack S. Hale and Christina Phung for proofreading the manuscript.

\section*{Conflict of interest statement}
The authors of this work have no conflicts of interest to disclose.

\end{document}